\documentstyle[12pt]{article}
\renewcommand{\baselinestretch}{1.3}
 1

\catcode`\@=11 
\def\lsim{\mathrel{\mathpalette\@versim<}}
\def\gsim{\mathrel{\mathpalette\@versim>}}
\def\@versim#1#2{\vcenter{\offinterlineskip
        \ialign{$\m@th#1\hfil##\hfil$\crcr#2\crcr\sim\crcr } }}
\makeatletter
\def\@seccntformat#1{\csname the#1\endcsname.\hskip 1em}
\makeatother

\begin{document}
\newcommand{\aven}{{\overline n}}
\newcommand{\avem}{{\overline m}}
\newcommand{\y}{{\it y}}
\newcommand{\as}{\alpha_s}
\newcommand{\A}{{\cal A}}
\newcommand{\B}{{\cal B}}
\newcommand{\Oa}{{\cal O}(\alpha_s)}
\newcommand{\Oaa}{{\cal O}(\alpha_s^2)}
\newcommand{\Oaaa}{{\cal O}(\alpha_s^3)}
\newcommand{\lam}{\Lambda_{\overline{MS}}}
\newcommand{\Rm}{{\it R-matching}}
\newcommand{\mRm}{{\it modified R-matching}}
\newcommand{\lnRm}{{\it lnR-matching}}
\newcommand{\mlnRm}{{\it modified lnR-matching}}
\newcommand{\aspi}{\tilde{\alpha}_s}
\newcommand{\asz}{\alpha_s(M_Z^2)}
 
\thispagestyle{empty}
\begin{flushright}
{\footnotesize\renewcommand{\baselinestretch}{.75}
  SLAC--PUB--7172\\
June 1996\\
}
\end{flushright}

\begin{center}
 {\Large \bf Measurement of the Charged Multiplicities in $b$, $c$ and 
          Light Quark Events from $Z^0$ Decays$^*$}

\vspace {1.5cm}

 {\bf The SLD Collaboration$^{**}$}\\
Stanford Linear Accelerator Center \\
Stanford University, Stanford, CA~94309

\vspace{4.0cm}

{\it To appear in Physics Letters B}

\end{center} 

\vfill

\normalsize
\vbox{
\uchyph=200
\brokenpenalty=200
\pretolerance=10000
\tolerance=2000
\nobreak
\penalty 5000
\hyphenpenalty=5000
\exhyphenpenalty=5000
\footnotesize\renewcommand{\baselinestretch}{1}\noindent
$^*$ This work was supported by Department of Energy
  contracts:
  DE-FG02-91ER40676 (BU),
  DE-FG03-91ER40618 (UCSB),
  DE-FG03-92ER40689 (UCSC),
  DE-FG03-93ER40788 (CSU),
  DE-FG02-91ER40672 (Colorado),
  DE-FG02-91ER40677 (Illinois),
  DE-AC03-76SF00098 (LBL),
  DE-FG02-92ER40715 (Massachusetts),
  DE-AC02-76ER03069 (MIT),
  DE-FG06-85ER40224 (Oregon),
  DE-AC03-76SF00515 (SLAC),
  DE-FG05-91ER40627 (Tennessee),
  DE-FG02-95ER40896 (Wisconsin),
  DE-FG02-92ER40704 (Yale);
  National Science Foundation grants:
  PHY-91-13428 (UCSC),
  PHY-89-21320 (Columbia),
  PHY-92-04239 (Cincinnati),
  PHY-88-17930 (Rutgers),
  PHY-88-19316 (Vanderbilt),
  PHY-92-03212 (Washington);
  the UK Science and Engineering Research Council
  (Brunel and RAL);
  the Istituto Nazionale di Fisica Nucleare of Italy
  (Bologna, Ferrara, Frascati, Pisa, Padova, Perugia);
  and the Japan-US Cooperative Research Project on High Energy Physics
  (Nagoya, Tohoku).}
\newpage

{\textwidth 15.0cm
\textheight 23.0cm
\oddsidemargin -.1cm
\evensidemargin -.1cm
\topmargin -1.7cm

\bigskip
\begin{center}
\bigskip
%
%
%
  \def\iADEL{$^{(1)}$}
  \def\iBOL{$^{(2)}$}
  \def\iBU{$^{(3)}$}
  \def\iBRUN{$^{(4)}$}
  \def\iUCSB{$^{(5)}$}
  \def\iUCSC{$^{(6)}$}
  \def\iCIN{$^{(7)}$}
  \def\iCSU{$^{(8)}$}
  \def\iCOLO{$^{(9)}$}
  \def\iCOL{$^{(10)}$}
  \def\iFER{$^{(11)}$}
  \def\iFRA{$^{(12)}$}
  \def\iILL{$^{(13)}$}
  \def\iLBL{$^{(14)}$}
  \def\iMIT{$^{(15)}$}
  \def\iMASS{$^{(16)}$}
  \def\iMISS{$^{(17)}$}
  \def\iMOSC{$^{(18)}$}
  \def\iNAG{$^{(19)}$}
  \def\iOREG{$^{(20)}$}
  \def\iPAD{$^{(21)}$}
  \def\iPERU{$^{(22)}$}
  \def\iPISA{$^{(23)}$}
  \def\iRUT{$^{(24)}$}
  \def\iRAL{$^{(25)}$}
  \def\iSOGANG{$^{(26)}$}
  \def\iSLAC{$^{(27)}$}
  \def\iTENN{$^{(28)}$}
  \def\iTOH{$^{(29)}$}
  \def\iVAND{$^{(30)}$}
  \def\iWASH{$^{(31)}$}
  \def\iWISC{$^{(32)}$}
  \def\iYALE{$^{(33)}$}
  \def\dead{$^{\dag}$}
  \def\andgen{$^{(a)}$}
  \def\andper{$^{(b)}$}
%
%
$^{**}$
\mbox{K. Abe                 \unskip,\iNAG}
\mbox{K. Abe                 \unskip,\iTOH}
\mbox{I. Abt                 \unskip,\iILL}
\mbox{T. Akagi               \unskip,\iSLAC}
\mbox{N.J. Allen             \unskip,\iBRUN}
\mbox{W.W. Ash               \unskip,\iSLAC$^\dagger$}
\mbox{D. Aston               \unskip,\iSLAC}
\mbox{K.G. Baird             \unskip,\iRUT}
\mbox{C. Baltay              \unskip,\iYALE}
\mbox{H.R. Band              \unskip,\iWISC
% ...
}
\it
  \vskip \baselineskip                   
  \vskip \baselineskip                   
%
%
%
  \iADEL
     Adelphi University,
     Garden City, New York 11530 \break
  \iBOL
     INFN Sezione di Bologna,
     I-40126 Bologna, Italy \break
  \iBU
     Boston University,
     Boston, Massachusetts 02215 \break
  \iBRUN
     Brunel University,
     Uxbridge, Middlesex UB8 3PH, United Kingdom \break
  \iUCSB
     University of California at Santa Barbara,
     Santa Barbara, California 93106 \break
  \iUCSC
     University of California at Santa Cruz,
     Santa Cruz, California 95064 \break
  \iCIN
     University of Cincinnati,
     Cincinnati, Ohio 45221 \break
  \iCSU
     Colorado State University,
     Fort Collins, Colorado 80523 \break
  \iCOLO
     University of Colorado,
     Boulder, Colorado 80309 \break
  \iCOL
     Columbia University,
     New York, New York 10027 \break
  \iFER
     INFN Sezione di Ferrara and Universit\`a di Ferrara,
     I-44100 Ferrara, Italy \break
  \iFRA
     INFN  Lab. Nazionali di Frascati,
     I-00044 Frascati, Italy \break
  \iILL
     University of Illinois,
     Urbana, Illinois 61801 \break
  \iLBL
     Lawrence Berkeley Laboratory, University of California,
     Berkeley, California 94720 \break
  \iMIT
     Massachusetts Institute of Technology,
     Cambridge, Massachusetts 02139 \break
  \iMASS
     University of Massachusetts,
     Amherst, Massachusetts 01003 \break
  \iMISS
     University of Mississippi,
     University, Mississippi  38677 \break
  \iNAG
     Nagoya University,
     Chikusa-ku, Nagoya 464 Japan  \break
  \iOREG
     University of Oregon,
     Eugene, Oregon 97403 \break
  \iPAD
     INFN Sezione di Padova and Universit\`a di Padova,
     I-35100 Padova, Italy \break
  \iPERU
     INFN Sezione di Perugia and Universit\`a di Perugia,
     I-06100 Perugia, Italy \break
  \iPISA
     INFN Sezione di Pisa and Universit\`a di Pisa,
     I-56100 Pisa, Italy \break
  \iRUT
     Rutgers University,
     Piscataway, New Jersey 08855 \break
  \iRAL
     Rutherford Appleton Laboratory,
     Chilton, Didcot, Oxon OX11 0QX United Kingdom \break
  \iSOGANG
     Sogang University,
     Seoul, Korea \break
  \iSLAC
     Stanford Linear Accelerator Center, Stanford University,
     Stanford, California 94309 \break
  \iTENN
     University of Tennessee,
     Knoxville, Tennessee 37996 \break
  \iTOH
     Tohoku University,
     Sendai 980 Japan \break
  \iVAND
     Vanderbilt University,
     Nashville, Tennessee 37235 \break
  \iWASH
     University of Washington,
     Seattle, Washington 98195 \break
  \iWISC
     University of Wisconsin,
     Madison, Wisconsin 53706 \break
  \iYALE
     Yale University,
     New Haven, Connecticut 06511 \break
  \dead
     Deceased \break
  \andgen
     Also at the Universit\`a di Genova \break
  \andper
     Also at the Universit\`a di Perugia \break
\rm
%
 
\end{center}
}

 
\vfill
\eject



\begin{center}
{\bf ABSTRACT }
\end{center}
\noindent
Average charged
multiplicities have been measured separately in $b$, $c$ and
light quark ($u,d,s$) events from $Z^0$ decays measured in the SLD
experiment. 
Impact parameters of charged tracks were used
to select enriched samples of $b$ and light quark events, 
and reconstructed charmed
mesons were used to select $c$ quark events. 
We measured the charged multiplicities:
$\aven_{uds} = 20.21 	\pm 0.10\hspace{1mm}(\rm{stat.}) 
			\pm 0.22\hspace{1mm}(\rm{syst.})$,
$\aven_{c}   = 21.28 	\pm 0.46\hspace{1mm}(\rm{stat.}) 
			^{+0.41}_{-0.36}\hspace{1mm}(\rm{syst.})$ 
and 
$\aven_{b}   = 23.14 	\pm 0.10\hspace{1mm}(\rm{stat.}) 
			^{+0.38}_{-0.37}\hspace{1mm}(\rm{syst.})$, 
from which
we derived 
the differences between the total
average charged multiplicities of $c$ or $b$ quark
events and light quark events:
$\Delta \aven_c  = 1.07 \pm 0.47\hspace{1mm}(\rm{stat.}) 
			^{+0.36}_{-0.30}\hspace{1mm}(\rm{syst.})$ and 
$\Delta \aven_b  = 2.93 \pm 0.14\hspace{1mm}(\rm{stat.}) 
			^{+0.30}_{-0.29}\hspace{1mm}(\rm{syst.})$. 
We compared these measurements with those at lower center-of-mass energies and
with perturbative QCD predictions. 
These combined results are in agreement with the 
QCD expectations and disfavor the hypothesis of
flavor-independent fragmentation.

\section{Introduction}
Heavy quark ($Q$=$c$,$b$) systems provide important laboratories for
experimental tests of the
theory of strong interactions, quantum chromodynamics (QCD). Since
the large quark mass $M_Q$ acts as a cutoff for soft gluon radiation,
some properties of these systems can be calculated accurately in
perturbative QCD. In other cases, however, where QCD calculations
assume massless quarks, the
products of heavy hadron decays can complicate the comparison of data
with the predictions for massless partons. It is therefore
desirable to measure properties of both light- and heavy-quark systems
as accurately as possible.
 
In this paper we consider one of the most basic observable properties of
high energy particle interactions, the multiplicity of charged
hadrons produced in the final state. We consider
hadronic $Z^0$ decays, which are believed to proceed via creation of a primary
quark-antiquark pair,
$Z^0$ $\rightarrow$ $q\bar{q}$,
which subsequently undergoes a fragmentation
process to produce the observed jets of hadrons. If the primary
event flavor $q$ can be identified experimentally, one can measure
the average charged multiplicity $\aven_q$ in events of that flavor,
for example $q=b,c,uds$, where $uds$ denotes the average over events of the
types $Z^0$ $\rightarrow$ $u\bar{u}$, $d\bar{d}$, and $s\bar{s}$. 
These are not only important properties of $Z^0$ decays, but,
if the average decay multiplicity of the {\it leading}
hadrons that contain the primary heavy quark or antiquark is subtracted from
$\aven_Q$ to yield the average {\it non-leading} multiplicity, 
can also
be used to test our understanding of the quark fragmentation process and its
dependence on the quark mass. 
The hypothesis of flavor-independent
fragmentation \cite{rowson,kisselev} implies
that this non-leading multiplicity in 
$e^+e^- \to Q\bar{Q}$ (``heavy quark'') events
at center-of-mass (c.m.) energy $W$
should be equal to the total multiplicity in e$^+$e$^- \rightarrow
u\bar{u}, d\bar{d}$, and $s\bar{s}$ (``light quark'') events at a
lower c.m. energy given by the average energy of the non-leading system,
$E_{nl}$ = $(1-\langle x_{E_Q}\rangle )W$, where 
$\langle x_{E_Q}\rangle$ = $2\langle E_Q\rangle/W$ 
is the mean fraction of
the beam energy carried by a heavy hadron of flavor $Q$.
 
Perturbative QCD predictions have been made \cite{schumm}
of the multiplicity {\it difference} between heavy- and
light-quark events, $\Delta\aven_Q$ = $\aven_Q-\aven_{uds}$.
In this case the suppression of soft gluon radiation 
caused by the heavy quark
mass leads to a depletion of the non-leading multiplicity, and
results in the striking prediction that $\Delta\aven_Q$ is independent of $W$
at the level of $\pm$0.1 tracks.
Numerical predictions of $\Delta\aven_b=5.5\pm1.3$
and $\Delta\aven_c=1.7\pm1.1$ were also given \cite{schumm}.
More recently,  improved calculations have been performed \cite{petrov},
confirming that the
energy-dependence is expected to be very small and predicting
$\Delta\aven_b$=3.53$\pm$0.23 and $\Delta\aven_c$=1.02$\pm$0.24 at $W=M_{Z^0}$.
 
In our previous paper \cite{nb} we measured $\aven_b$ and
$\Delta\aven_b$ using the sample
of about 10,000 hadronic $Z^0$ decays recorded by the SLD experiment
in the 1992 run. By comparing with similar
measurements at lower c.m. energies \cite{rowson,delco,tpc,tasso} 
we found that $\Delta\aven_b$
was consistent with an energy-independent value, and in agreement with
the prediction of \cite{schumm}.
This result was subsequently confirmed by the 
DELPHI \cite{delphi} and OPAL \cite{opal} Collaborations. 
The dominant uncertainty in our
measurement resulted from lack of knowledge of the charged multiplicity
in $Z^0$ $\rightarrow$ $c\bar{c}$ events, $\aven_c$. 
In this paper we present simultaneous
measurements of $\aven_b$, $\aven_c$ and $\aven_{uds}$
based upon the sample of
about 160,000 hadronic $Z^0$ decays collected by SLD between 1992 and
1995, and using the SLD micro-vertex detector and tracking system
for flavor separation.
By measuring $\aven_c$ and $\aven_{uds}$ directly we have reduced
the systematic uncertainty on $\Delta\aven_b$ substantially, and have also
derived $\Delta\aven_c$, which allows us to compare with the QCD
predictions for the charm system
and with the only other measurement of this
quantity \cite{opal} at the $Z^0$ resonance.
This measurement supersedes our previous measurements of $\aven_b$ and 
$\Delta\aven_b$ \cite{nb}.

\section{Apparatus and Hadronic Event Selection}
The e$^+$e$^-$ annihilation events produced at the $Z^0$ resonance
by the SLAC Linear Collider (SLC)
were recorded using the SLC Large Detector (SLD).
A general description of the SLD can be found elsewhere \cite{sld}.
The trigger and selection criteria for isolating 
hadronic $Z^0$ boson decays are
described elsewhere \cite{alphas}.
 
The analysis presented here used the charged tracks measured 
in the central drift chamber (CDC) \cite{cdc} and 
in the vertex detector (VXD) \cite{vxd}.
A set of cuts was applied to the data to select well-measured tracks,
which were used for multiplicity counting, 
and events well-contained within the detector acceptance.
The well-measured tracks were required to have
(i) a closest approach transverse to the beam axis within 5 cm,
and within 10 cm along the axis from the measured interaction point; 
(ii) a polar angle $\theta$ with respect to the beam axis within 
$\mid \cos \theta \mid < 0.80$; and
(iii)~a momentum transverse to the beam axis $p_{\bot} > 0.15$ GeV/c.
Events were required to have
(i) a minimum of seven such tracks;
(ii) a thrust axis \cite{thrust} direction 
within $\mid \cos \theta_T \mid < 0.71$; and
(iii)~a total visible energy $E_{vis}$ of at least 20~GeV,
which was calculated from the selected tracks assigned the charged pion mass;
114,499 events passed these cuts.
The background in the selected event sample was estimated to be $0.1\pm 0.1\%$,
dominated by $Z^0 \rightarrow \tau^+ \tau^-$ events. 

While the multiplicity measurement relied primarily on information
from the CDC, 
the additional information from the VXD provided the more
accurate impact parameter measurement, and 
$D$ meson vertex reconstruction, used for selecting samples
enriched in light ($u$,$d$,$s$) and $b$ events, and $c$ events, respectively.
In addition to the requirements for well-measured tracks, 
``impact parameter quality'' tracks were required to have
(i) at least one VXD hit; 
(ii) a closest approach transverse to the beam axis within 0.3 cm, and 
within 1.5 cm along the axis from the measured interaction point;
(iii) at least 40 CDC hits, with the first hit at a radius 
less than 39 cm; 
(iv) an error on the impact parameter 
transverse to the beam axis less than 250 $\mu$m;  
and
(v) a fit quality of the combined CDC+VXD track $\chi^2/d.o.f<5$.
We also removed tracks from candidate $K_s^0$ and $\Lambda$
decays and $\gamma$-conversions found by kinematic reconstruction of 
two-track vertices.

All impact parameters used in this analysis were for tracks
projected into the ($x-y$) plane perpendicular to the beam axis, 
and were measured
with respect to an average primary vertex.
The average primary vertex was derived from fits to $\sim$30 
sequential hadronic events close in time to the event under study,
with a measured precision of $\sigma_{PV}=(7\pm2)\mu$m \cite{rb}.
The impact parameter $\delta$ was derived by applying a sign to the 
distance of closest approach such that $\delta$ is positive when
the vector from the primary vertex to the point at which the track 
intersects the thrust axis makes an acute angle with respect to
the track direction.
Including the uncertainty on the average primary vertex the measured
impact parameter uncertainty $\sigma_\delta$ for the overall
tracking system approaches 11 $\mu$m for high momentum tracks,
and is 76 $\mu$m at $p_{\bot}\sqrt{\sin\theta}$=1 GeV/c \cite{rb}. 

\section{Selection of Flavor-Tagged Samples}
We divided each event into two hemispheres separated by the plane
perpendicular to the thrust axis.
We then applied three flavor tags to each hemisphere. 
In order to reduce potential tagging bias we measured the average charged 
multiplicity in hemispheres opposite those tagged.
Impact parameters of charged tracks were used
to select enriched samples of $b$ or light quark hemispheres, 
and reconstructed charmed
mesons were used to select $c$ quark hemispheres.

In each hemisphere
we counted the number of impact parameter quality
tracks $n_{sig}$ that had an impact parameter significance of 
$\delta_{norm} = \delta / \sigma_\delta>$3.0.
Fig. 1 shows the distribution of $n_{sig}$ upon which is 
superimposed a Monte Carlo simulated 
distribution in which the flavor composition is shown.
For our Monte Carlo study we used the JETSET 7.4 event generator
\cite{jetset74} with parameter values tuned to hadronic $e^+e^-$ 
annihilation data \cite{tune}, combined with 
a simulation of $B$-decays tuned to $\Upsilon$(4S) data \cite{rb}, 
and a simulation of the SLD.
A more detailed discussion of flavor tagging using impact parameters 
can be found in \cite{rb}.
The Monte Carlo simulation
reproduces the data well and shows that most light quark
hemispheres have $n_{sig}$=0 and that the $n_{sig}\geq$3 region is
dominated by $b$ quark hemispheres.
Hemispheres were tagged as light or $b$ quark by requiring $n_{sig}=0$
or $n_{sig}\ge 3$, respectively.
Table 1 shows the number of light and $b$ quark
tagged hemispheres and their flavor compositions
estimated from the simulation.
%
\begin{table}[t]
 
\begin{center}
\footnotesize
\renewcommand{\arraystretch}{1.0}
\begin{tabular}{|c|c|c|c|c|}  \hline
\multicolumn{2}{|c|}{}   & $uds$-tag & $c$-tag & $b$-tag \\ \hline
\multicolumn{2}{|c|}{\# hemispheres} 
            & 154,151  &  976  & 9,480 \\ \hline
            & $uds$ & 0.752$\pm$0.001 $\pm$0.004 & 0.074$\pm$0.002 $\pm$0.014
                   & 0.014$\pm$0.001 $\pm$0.001 \\ 
		\cline{2-5} 
composition &  $c$ & 0.158$\pm$0.001 $\pm$0.006 & 0.640$\pm$0.008 $\pm$0.025
                   & 0.048$\pm$0.001 $\pm$0.005 \\ 
		\cline{2-5}
            &  $b$ & 0.089$\pm$0.001 $\pm$0.004 & 0.286$\pm$0.005 $\pm$0.022
                   & 0.938$\pm$0.001 $\pm$0.006 \\ \hline
\end{tabular}
\end{center}
\footnotesize
Table 1.
Numbers of hemispheres and fractional compositions of $uds$, $c$ and $b$ 
quarks in the tagged hemispheres. 
The first
quoted errors represent the errors due to the limited size of the
Monte Carlo sample
and the second are due to the uncertainties from 
the modelling of heavy hadron production and decay. 
\end{table}
\normalsize

From Fig. 1 it is clear that
an impact parameter tag does not provide a high-purity 
sample of $c$ quark hemispheres.
For this purpose we required at least one prompt 
$D^{*+}$ or $D^+$ 
meson\footnote{In this paper charge-conjugate cases are always implied.}
reconstructed in a hemisphere.
This tag is similar to that described in \cite{ac}. 
The $D^{*+}$ mesons were identified using the decay 
$D^{*+} \to \pi_s^+ D^0$, where 
$\pi_s^+$ is a low-momentum pion and 
the $D^0$ decays via 
$D^0 \to K^-\pi^+$ (``three-prong''), 
$D^0 \to K^-\pi^+\pi^0$ (``satellite''),
or $D^0 \to K^-\pi^+\pi^+\pi^-$ (``five-prong'') modes.
The $D^+$ mesons were indentified using the decay mode 
$D^+ \to K^-\pi^+\pi^+$.
$D$ meson candidates were formed from all combinations of well-measured
tracks with at least one VXD hit.
$D^0$ candidates were formed by combining
two (for the three-prong and satellite modes) 
or four (for the five-prong mode) charged tracks with zero net charge, 
and by assigning the $K^-$
mass to one of the particles and $\pi^+$ mass to the others. 

For $D^{*+}$ candidates, we first required 
a candidate $D^0$ in the mass range
1.765 GeV/c$^2 < M_{D^0}^{cand.} <$1.965 GeV/c$^2$ 
(three-prong),  
1.815 GeV/c$^2 < M_{D^0}^{cand.} <$1.915 GeV/c$^2$ 
(five-prong), or 
1.500 GeV/c$^2 < M_{D^0}^{cand.} <$1.700 GeV/c$^2$ 
(satellite). 
$D^{*+}$ candidates were then required to pass either a set of kinematic
cuts or a set of decay length cuts to suppress combinatorial backgrounds
and backgrounds from $B \to D^{*+}$ decays.
The kinematic cuts are: 
(i) $|\cos\theta_{KD^0}|<0.9$ (three-prong and satellite modes)
    and 
    $|\cos\theta_{KD^0}|<0.8$ (five-prong mode),
    where $\theta_{KD^0}$ is the angle between the 
    $D^0$ direction in the 
    laboratory frame and the $K$ direction in the $D^0$ rest frame,
(ii) $p_{\pi^+_s}>$1 GeV/c, and  
(iii) $x_{E_{D^{*+}}}>$0.4 for the three-prong and satellite modes
       and 
      $x_{E_{D^{*+}}}>$0.5 for the five-prong mode,
where $x_{E_{D^{*+}}}=2E_{D^{*+}}/W$ and $E_{D^{*+}}$ is the $D^{*+}$ energy.
For the decay length analysis we performed a fit of the $D^0$ tracks to a
common vertex and 
calculated the decay length, $L^0$, between the primary vertex and
this $D^0$ decay vertex, and its error, $\sigma_{L^0}$.  
The decay length cuts are:
(i) a $\chi^2$ probability$>$1\% for the vertex fit
to the $D^0$ tracks, 
(ii) a decay length significance $L^0/\sigma_{L^0}>$2.5,
(iii) the two-dimensional impact parameter of the $D^0$ momentum
vector to the interaction point $<$20$\mu$m, and
(iv) $x_{E_{D^{*+}}}>$0.2 for the three-prong and satellite modes
      and
     $x_{E_{D^{*+}}}>$0.4 for the five-prong mode.

For all $D^{*+}$ candidates
we required the proper decay time of the $D^0$, 
$\tau_{\rm proper}=L^0/\beta\sqrt{1-\beta^2}$, where $\beta=p_{D^0}/E_{D^0}$
and
$p_{D^0}$ and $E_{D^0}$ are the reconstructed momentum and energy, 
respectively, of the candidate $D^0$ meson, 
to be in the range 0$<\tau_{\rm proper}<$1ps.
Figs. 2(a), (b) and (c) show the distribution of $\Delta M$, 
where $\Delta M \equiv M_{D^{*+}}^{cand}-M_{D^0}^{cand}$,
after the above cuts for the 
three $D^0$ decay modes,
upon which is superimposed the Monte Carlo simulated distribution in which 
the flavor composition is 
shown\footnote{In the Monte Carlo simulation 
the production cross section and branching fractions, and
normalization of the $\Delta M$ distributions in the region
$\Delta M>0.15$ GeV/c$^2$, for 
the $D^0\to K^-\pi^+$ and $D^0\to K^-\pi^+\pi^0$ modes were
adjusted to match the data in Fig. 2, as described in Ref. [19].
The adjustment was small and included in the systematic errors.}.
A hemisphere was tagged as $c$ if it contained a $D^{*+}$ candidate with
$\Delta M<$0.15 GeV/c$^2$.  

$D^+ \to K^-\pi^+\pi^+$ candidates were formed by combining two
tracks of the same sign with one track of the opposite sign, where
all three tracks were required to have momentum $p>$1 GeV/$c$.
The two like-sign tracks were assigned $\pi^+$ masses, 
the opposite-sign track was assigned the $K^-$ mass, and
all three tracks were fitted to a common vertex. 
A series of cuts was applied to reject random combinatoric, $D^{*+}$, and
B-decay backgrounds. 
We required:
(i) $x_{E_{D^{+}}}>0.4$,
(ii) $\cos\theta_{KD^+}>-0.8$,
where $\theta_{KD^+}$ is the
angle between the directions of the $D^+$ in the laboratory frame and
the $K^-$ in the $D^+$ rest frame, 
(iii) the mass differences  
$M(K^-\pi^+\pi^+)-M(K^-\pi^+)$ for each of the two
pions to be greater than 0.16 GeV/c$^2$, 
(iv) the normalized $D^+$ decay length $L^+/\sigma_{L^+}>$3.0, and 
(v) the projection of the angle between the $D^+$ momentum vector and the
vertex flight direction to be less than 5 mrad in the plane perpendicular
to the beam axis and
less than 20 mrad in the plane containing the beam axis.
A hemisphere was $c$-tagged if it contained a 
$D^+ \to K^-\pi^+\pi^+$ candidate
in the mass range
1.800 GeV/c$^2<M(K^-\pi^+\pi^+)<$1.940 GeV/c$^2$.
Fig. 2(d) shows the mass $M(K^-\pi^+\pi^+)$ distribution of the 
data upon which is 
superimposed the Monte Carlo simulated distribution
in which the flavor composition is shown. 

The union of the three samples of $D^{*+}$ candidates and 
the sample of $D^+$ candidates was used to tag
$c$ quark hemispheres.
The flavor composition of these tagged hemispheres is shown in Table 1.
Approximately 400 of the $c$-tagged hemispheres were also
tagged as either $b$ or $uds$ hemispheres. 
Monte Carlo studies indicated that these were mostly true $c$-hemispheres.
The exclusion of
these hemispheres from the $b$- and $uds$-tagged samples 
was found to have negligible effect on the final
results.

\section{Measurement of Charged Multiplicities}
Well-measured charged tracks defined in Section 2 were
counted in the hemispheres opposite those tagged. 
The measured average hemisphere multiplicities $\avem_i$ 
($i=uds,c,b$) were  
$\avem_{uds}=8.94\pm0.01$, $\avem_c=9.15\pm0.12$ and
$\avem_b=9.99\pm0.04$ (statistical errors only). 

The $\avem_i$ are related to the true average
multiplicities $\aven_j$ ($j=uds,c,b$) of $uds$, $c$ and $b$ quark events by:
\begin{eqnarray}
2\times\avem_{i}  = P_{i,uds} C_{i,uds} \aven_{uds} 
       + P_{i,c} ( C_{i,c}^{dk} \aven_c^{dk} + C_{i,c}^{nl} \aven_c^{nl} )
       + P_{i,b} ( C_{i,b}^{dk} \aven_b^{dk} + C_{i,b}^{nl} \aven_b^{nl} )  
\end{eqnarray} 
where: 
$P_{i,j}$ is the fraction of hemispheres of quark type $j$ 
in the $i$-tagged hemisphere sample; 
$\aven_j=\aven^{dk}_j+\aven^{nl}_j$ ($j\not= uds$), and 
$\aven^{dk}_j$ is the true average multiplicity originating from 
the decay products of $j$-hadrons and 
$\aven^{nl}_j$ is that originating from the non-leading particles;
$C_{i,uds}$ is the ratio of the average number of measured charged tracks
in light quark  
hemispheres opposite $i$-tagged hemispheres, 
to the average number of 
charged tracks in true light quark hemispheres; 
$C_{i,j}^{dk}$ ($j \neq uds$) 
is the ratio of the average number of 
measured charged tracks originating from the 
decay products of $j$-hadrons\footnote{We include the products 
of both strongly and weakly decaying heavy hadrons.}
in hemispheres opposite
$i$-tagged hemispheres, to the average number of tracks 
originating from the decay products of $j$-hadrons;
$C_{i,j}^{nl}$ ($j \neq uds$) 
is the ratio of the average number of 
measured charged tracks originating from the
non-leading particles in true $j$-quark
hemispheres opposite those tagged as $i$-quark hemispheres,
to the average number of 
tracks originating from non-leading particles in 
true $j$-quark hemispheres.
The constants $P$ are shown in Table 1.
The constants $C$ were also calculated from our Monte Carlo simulation
and are shown in Table 2; they
account for the effects of detector acceptance and 
inefficiencies, for tracks from beam-related backgrounds and interactions
in the detector material,
and for biases introduced by the event and tagged-sample
selection criteria. 
We included in the generated multiplicity any prompt charged track
with mean lifetime greater than $3\times 10^{-10}$s,
or any charged 
decay product with mean lifetime greater than $3\times 10^{-10}$s
of a particle with mean lifetime less than $3\times 10^{-10}$s.

We fixed
$\aven_c^{dk}$=5.20 and 
$\aven_b^{dk}$=11.10, using the measured values from
\cite{markiii,cleo,argus}
with the addition of 0.20 and 0.22 tracks, respectively,
estimated from the Monte Carlo simulation,
to account for the effects of higher mass states of heavy hadrons
produced in $Z^0$ decays.
\begin{table}[t]
 
\begin{center}
\footnotesize
\renewcommand{\arraystretch}{1.0}
\begin{tabular}{|c|c|c|c|c|c|}  \hline
 $j$  & $uds$ & \multicolumn{2}{|c|}{$c$} & \multicolumn{2}{|c|}{$b$} 
\\ \cline{1-1} \cline{3-6}
 $i$  &       & $dk$ & $nl$               & $dk$ & $nl$  \\ \hline
$uds$ & 0.875$\pm$0.001$\pm$0.001
      	& 0.798$\pm$0.002$^{+0.006}_{-0.005}$
	& 0.885$\pm$0.002$^{+0.013}_{-0.015}$
	& 0.820$\pm$0.002$^{+0.024}_{-0.020}$
	& 0.887$\pm$0.003$^{+0.022}_{-0.021}$ \\ \hline
$c$   & 0.803$\pm$0.019$^{+0.005}_{-0.007}$
	& 0.831$\pm$0.011$\pm$0.004
	& 0.864$\pm$0.009$^{+0.014}_{-0.017}$
	& 0.854$\pm$0.013$^{+0.030}_{-0.024}$
	& 0.849$\pm$0.015$\pm$0.026 \\ \hline
$b$   & 0.875$\pm$0.015$^{+0.005}_{-0.002}$
	& 0.816$\pm$0.013$\pm$0.003
	& 0.887$\pm$0.010$^{+0.016}_{-0.018}$ 
	& 0.854$\pm$0.003$^{+0.025}_{-0.021}$
	& 0.893$\pm$0.004$\pm$0.025 \\ \hline
\end{tabular}
\end{center}
\footnotesize
Table 2.
The constants $C$ calculated from the Monte Carlo simulation. 
The first quoted errors are statistical and arise from the finite size of the
Monte Carlo sample.  
The second are due to the uncertainties from C and B hadron production
and decay.
\end{table}
\normalsize 

We then solved
eqns. (1) to obtain the average charged
multiplicities per event,
$\aven_{uds} = 20.21 \pm 0.10$,
$\aven_{c} = 21.28 \pm 0.46$ and 
$\aven_{b} = 23.14 \pm 0.10$ (statistical errors only).
The multiplicity
differences between $c$ and light quark events, and 
$b$ and light quark events are, respectively
\[\begin{array}{rl}
\Delta \aven_c  & = 1.07 \pm 0.47\hspace{1mm}(\rm{stat.})  \\
\Delta \aven_b  & = 2.93 \pm 0.14\hspace{1mm}(\rm{stat.}).  \\
\end{array}\]
   
\section{Systematic Errors}
Experimental systematic errors arise from uncertainties in 
modelling the acceptance, efficiency and resolution of the detector.
Systematic uncertainties also arise from errors on the experimental
measurements that function as the input parameters to the modelling
of the underlying physics processes, such as errors on the
modelling of $b$ and $c$ fragmentation and decays of B and C hadrons.

The effect of uncertainty in the tracking efficiency was
estimated to cause a common $\pm$0.9\% variation of the constants $C$.
The effect of uncertainty in the corrections for the residual
$\gamma$ conversions and fake tracks
was estimated to cause a common $\pm$0.5\% variation of the constants $C$.
Statistical effects from the limited Monte Carlo sample size were 
also considered.
These errors, summarized in Table 3, 
were added in quadrature to obtain a total systematic error
due to detector modelling.
Note that the uncertainties in total track reconstruction efficiency
are the dominant source of systematic error for $\aven_{uds}$
and $\aven_b$, but are small for the differences 
$\Delta\aven_c$ and $\Delta\aven_b$.
In the case of $\aven_c$, $\Delta\aven_c$ and $\Delta\aven_b$ 
the statistical error from the limited Monte Carlo sample
size is dominant.
\begin{table}[t]
\begin{center}
\footnotesize
\renewcommand{\arraystretch}{1.0}
\begin{tabular}{|c|ccccc|}  \hline
Source of Uncertainty  & $\aven_{uds}$ & $\aven_c$ 
		       & $\aven_b$     & $\Delta \aven_c$ 
		       & $\Delta \aven_b$ \\ \hline
Tracking efficiency    & $\pm$0.182 & $\pm$0.194 & $\pm$0.205
			& $\pm$0.012 & $\pm$0.023
\\
$\gamma$ conversion \& fake tracks
		& $\pm$0.101 & $\pm$0.108 & $\pm$0.114 
		& $\pm$0.007 & $\pm$0.013  
\\ 
Monte Carlo statistics & $\pm$0.046 & $\pm$0.212 & $\pm$0.045 
		& $\pm$0.217 & $\pm$0.064  \\ \hline
Total		& $\pm$0.213 & $\pm$0.307 
		& $\pm$0.239 & $\pm$0.217 
		& $\pm$0.069  \\ \hline
\end{tabular}
\end{center}
\footnotesize
Table 3.
Systematic errors due to detector modelling. 
\end{table}
\normalsize

We performed several consistency checks on our results.
We checked that our Monte Carlo simulation
showed good agreement with the data for track
$p_\bot$ and $\cos\theta$ 
distributions in the hemispheres opposite those tagged.
We then varied the thrust axis containment cut
within 0.5$\le |\cos\theta_T| \le $0.8.
To check for possible bias from our hemisphere tags
the cut on the track significance $\delta_{norm}$ was varied from
2.0 to 4.0 for the light and $b$ quark hemisphere tags, 
and $D^{*+}$ and $D^+$ mesons were 
considered separately as a $c$ quark hemisphere tag.
We also removed hemipheres tagged as both $c$ and $uds$ or $b$. 
Finally, we performed our analysis separately in 2- and $\geq3$-jet event
samples selected using the 
Durham algorithm \cite{durham} with $y_{cut}$=0.003, to check for
any possible bias in multi-jet events. 
In each case
all the re-evaluated $\aven_i$ were found to be consistent 
with our central values of $\aven_i$ within the statistical errors.

In order to estimate the systematic errors due to uncertainties in
modelling heavy hadron production and decay 
we used an event re-weighting scheme
to vary the multiplicity distributions in the Monte Carlo simulation 
and to obtain modified values of the constants $C$ and $P$.
The effect of uncertainty in heavy flavor fragmentation was estimated 
by varying the $\epsilon$ parameter of the 
Peterson fragmentation function \cite{peterson} 
used as input to generate the Monte Carlo sample,
corresponding to $\delta<x_E>$=$\pm$0.012 and $\pm$0.011 
for $c$ and $b$ quarks respectively, corresponding to the
average errors in measurements of these quantities \cite{epsilon}.
The average B hadron lifetime was varied by $\pm$0.1 ps for B mesons 
and $\pm$0.3 ps for B baryons \cite{pdb}.
The effect of varying the B baryon production rate in $b$ events by
$\pm$3\% \cite{rb} was also examined.
Absolute variations of $\pm$6\% and $\pm$4\% 
were applied to the $B\to D^+$ 
branching ratio and $c\to D^+$ branching ratio,
respectively \cite{rb}.
The effect of the present experimental uncertainties in the 
branching fractions, $R_c=\Gamma(Z^0\to c\bar{c})/\Gamma(Z^0\to q\bar{q})$ 
	and $R_b=\Gamma(Z^0\to b\bar{b})/\Gamma(Z^0\to q\bar{q})$, 
of $\delta R_c$=$\pm$0.020 and $\delta R_b$=$\pm$0.003
respectively
\cite{pdb}
were also included. 
The decay multiplicities of C and B hadrons were varied by
$\pm$0.26 and $\pm$0.36 charged tracks, respectively
\cite{markiii,cleo,argus}.
For the  $D^{*+}$ analysis
we also accounted for the adjustment of the production 
cross section and branching
fractions for the $D^0\to K^-\pi^+$ and $D^0\to K^-\pi^+\pi^0$ modes 
in the Monte Carlo by assigning the full shift of the Monte Carlo
simulated distribution as a systematic error.
These uncertainties, summarized in Table 4, were added in quadrature to 
obtain total systematic uncertainties due to C and B hadron modelling.
For $\Delta\aven_c$ ($\Delta\aven_b$) the dominant contributions were
from the uncertainties in $c$ ($b$) fragmentation and $R_c$ ($R_b$).
\begin{table}[t]
\begin{center}
\footnotesize
\renewcommand{\arraystretch}{1.0}
\begin{tabular}{|c|c|ccccc|}  \hline
Source of Uncertainty  & Variation 
		       & $\aven_{uds}$ & $\aven_c$ 
		       & $\aven_b$     & $\Delta\aven_c$ 
		       & $\Delta\aven_b$ \\ \hline
$b$ fragmentation      & $\langle x_{E_b} \rangle$=0.700$\pm$0.011 
			& $\pm$0.001 & $^{+0.002}_{-0}$ & $^{+0.288}_{-0.281}$
			& $^{+0.004}_{-0}$ & $^{+0.289}_{-0.279}$ \\ 
$B$ meson lifetime      & $\tau_b$=1.55$\pm$0.1 ps 
			& $\pm$0.001 & $^{+0.027}_{-0.028}$ 
			& $^{+0.010}_{-0.007}$ & $^{+0.027}_{-0.026}$ 
			& $^{+0.012}_{-0.007}$ \\ 
$B$ baryon lifetime     & $\tau_b$=1.10$\pm$0.3 ps 
			& $^{+0}_{-0.008}$ & $^{+0.032}_{-0.036}$ 
			& $^{+0.008}_{-0.001}$ & $^{+0.041}_{-0}$ 
			& $^{+0.012}_{-0}$ \\ 
$B$ baryon prod. rate   & $f_{\Lambda_b} = 9\% \pm 3\%$ 
			& $^{+0.004}_{-0.001}$ & $^{+0}_{-0.001}$ 
			& $^{+0.021}_{-0.020}$ & $^{+0.001}_{-0.003}$ 
			& $^{+0.019}_{-0.018}$ \\ 
$R_b$ ($b$ fraction)    & 0.221$\pm$0.003 
			& $\pm$0.001 & $^{+0.007}_{-0.006}$ 
			& $^{+0.041}_{-0.040}$ & $^{+0.007}_{-0.008}$ 
			& $^{+0.040}_{-0.039}$ \\ 
$B \to D^+ + X$ fraction & 0.17$\pm$0.06 
			& $^{+0}_{-0.007}$ & $^{+0.054}_{-0}$ 
			& $^{+0}_{-0.036}$ & $^{+0.053}_{-0}$
 			& $^{+0}_{-0.024}$ \\ 
$c$ fragmentation      & $\langle x_{E_c} \rangle$=0.494$\pm$0.012 
			& $^{+0.008}_{-0.010}$ & $^{+0.236}_{-0.155}$ 
			& $^{+0.004}_{-0.002}$ & $^{+0.244}_{-0.151}$
			& $^{+0.015}_{-0.008}$\\ 
$R_c$ ($c$ fraction)    & 0.171$\pm$0.020 
			& $^{+0.026}_{-0.027}$ & $^{+0.081}_{-0.099}$ 
			& $^{+0.006}_{-0.007}$ & $^{+0.107}_{-0.126}$
 			& $\pm$0.033 \\ 
$c\bar{c} \to D^+ + X$ fraction & 0.20$\pm$0.04 
			& $^{+0.004}_{-0.003}$ & $^{+0.035}_{-0.039}$ 
			& $\pm$0.006 & $^{+0.039}_{-0.042}$ 
			& $^{+0.010}_{-0.009}$ \\ \hline
$\aven_c^{dk}$	        & 5.20$\pm$0.26 
			& $\pm$0.003 & $^{+0.010}_{-0.009}$ 
			& $^{+0.001}_{-0}$ 
			& $^{+0.005}_{-0.006}$ & $\pm$0.003      \\ 
$\aven_b^{dk}$	        & 11.10$\pm$0.36 
			& $\pm$0.003 & $^{+0.009}_{-0.008}$ & $\pm$0.016  
			& $\pm$0.012 & $\pm$0.013      \\ \hline
$D^0 \to K^-\pi^+$, $D^0\to K^-\pi^+\pi^0$ production	        
			& $-$20\% & $-$0.013 & +0.062 & $-$0.003  
			& +0.075 & +0.010      \\ \hline
Total	                & 
			& $^{+0.028}_{-0.034}$ & $^{+0.269}_{-0.194}$ 
			& $^{+0.293}_{-0.287}$ & $^{+0.289}_{-0.203}$
			& $^{+0.296}_{-0.286}$ \\ \hline 
\end{tabular}
\end{center}
\footnotesize
Table 4.
Systematic uncertainties due to heavy hadron modelling. 
\end{table}
\normalsize
\vspace{5mm}
 
\section{Summary and Conclusions}
Combining systematic uncertainties in quadrature we obtain:
\[\begin{array}{rl}
\aven_{uds} = & 20.21 \pm 0.10\hspace{1mm}(\rm{stat.}) 
		      \pm 0.22\hspace{1mm}(\rm{syst.}) \\
\aven_{c}   = & 21.28 \pm 0.46\hspace{1mm}(\rm{stat.}) \hspace{1mm}
	^{\textstyle +\hspace{0.8mm}0.41}
	_{\textstyle -\hspace{0.8mm}0.36}\hspace{0.8mm}(\rm{syst.}) 
			\hspace{1cm}  \\
\aven_{b}   = & 23.14 \pm 0.10\hspace{1mm}(\rm{stat.}) \hspace{1mm} 
	^{\textstyle +\hspace{0.8mm}0.38}
	_{\textstyle -\hspace{0.8mm}0.37}\hspace{0.8mm}(\rm{syst.}). \\
\end{array}\] 
Subtracting $\aven^{dk}_c$=5.20 and $\aven^{dk}_b$=11.10 from our
measured $\aven_c$ and $\aven_b$ respectively,
we obtained the average non-leading multiplicities 
$\aven^{nl}_c$ = 16.08 $\pm$ 0.46(stat.) $^{+0.41}_{-0.36}$(syst.) 
and 	
$\aven^{nl}_b$ = 12.04 $\pm$ 0.10(stat.) $^{+0.38}_{-0.37}$(syst.).
The hypothesis of flavor-independent fragmentation implies that
$\aven^{nl}_Q(W) =\aven_{uds}([1-\langle x_{E_Q}(W)\rangle ]W)$.
Fig. 3(a) shows 
our measurement of $\aven_{uds}$ plotted at $W=M_Z$, and our measurements of $\aven^{nl}_c$ and $\aven^{nl}_b$
plotted at the appropriately reduced non-leading energy
$[1-\langle x_{E_Q}(W)\rangle ]W$. 
Previous measurements of these quantities 
\cite{rowson,delco,tpc,tasso,delphi,opal,markii2,topaz}
are also shown.
The curve is a fit
to the energy dependence of the $\aven_{uds}$ measurements shown and those at 
5 $<W<$ 92 GeV \cite{markii2}.
Fig. 3(b) shows the differences between the non-leading data points
in Fig. 3(a) and the curve.
A linear fit to these differences (Fig. 3(b))
yields a slope of $s=1.14\pm0.32$ tracks/ln(GeV). 
This differs from the expectation for identical energy dependence, $s=0$, by 3.6 standard deviations, indicating that the hypothesis of flavor-independent fragmentation is disfavored at this level.  

Combining systematic uncertainties in quadrature we obtain:
\[\begin{array}{rl}
\Delta \aven_c  & = 1.07 \pm 0.47\hspace{1mm}(\rm{stat.}) 
	^{\textstyle +\hspace{0.8mm}0.36}
	_{\textstyle -\hspace{0.8mm}0.30}\hspace{0.8mm}(\rm{syst.}) \\
\Delta \aven_b  & = 2.93 \pm 0.14\hspace{1mm}(\rm{stat.}) 
	^{\textstyle +\hspace{0.8mm}0.30}
	_{\textstyle -\hspace{0.8mm}0.29}\hspace{0.8mm}(\rm{syst.}) .\\
\end{array}\] 
Fig. 4 shows our measurements of 
$\Delta \aven_c$ and $\Delta \aven_b$ 
together with those from other experiments
\cite{rowson,delco,tpc,tasso,delphi,opal,markii2},
at the respective c.m. energies.
The new result for $\Delta\aven_b$ is consistent with our previous measurement
\cite{nb} and with the measurements from LEP \cite{delphi,opal}
and Mark-II \cite{markii2}, and that for 
$\Delta\aven_c$ is consistent with the OPAL measurement \cite{opal}.
Linear fits to the $\Delta\aven_c$ and $\Delta\aven_b$ 
data as a function of $\ln(W)$ yield
slopes of $s$=$-$1.33$\pm$1.04 and $s$=$-$1.43$\pm$0.82 tracks/ln(GeV), 
respectively.
These slopes are consistent with the 
perturbative QCD prediction of energy independence [3],
$s$=0, at the level of 1.3 and 1.7 standard deviations, respectively.

Comparing our measurements of $\Delta\aven_c$ and $\Delta\aven_b$ with the
predictions of Refs. [3,4] 
(Fig. 4) we found that
both were in good agreement with the predictions of
Ref. [4], while the former was in good agreement
with the prediction of Ref. [3], and the latter within
1.7$\sigma$ of this prediction.
 
As a result of the accurate measurements of 
$\Delta\aven_c$ and $\Delta\aven_b$ at $W=M_{Z^0}$,
constraints on the energy dependence of these quantities
are now limited by the uncertainties in the lower energy
measurements. In order to improve the constraints on
the validity of perturbative QCD calculations at the scales
M$_b$ or M$_c$,
it is necessary to improve the accuracy
of the measurements of $\Delta\aven_b$ and $\Delta\aven_c$, respectively,
at lower energies, and/or
extend the $\ln(W)$ lever-arm of such measurements.
It would thus be desirable to have measurements of $\Delta\aven_c$
from the continuum below the $\Upsilon(4S)$, and for both $\Delta\aven_c$ and 
$\Delta\aven_b$ to be
measured at LEP-II and $e^+e^-$ colliders at even
higher energies.

 
\section*{Acknowledgements}
We thank the personnel of the SLAC accelerator department and the
technical
staffs of our collaborating institutions for their outstanding efforts
on our behalf.

\pagebreak 

\vfill
\eject
  
\clearpage
\section*{Figure captions }
 
\noindent
{\bf Figure 1}.
The distribution of the 
number of tracks per hemisphere $n_{sig}$ that miss the interaction
point by more
than 3$\sigma$ in the $x$-$y$ plane. 
The points represent the data distribution and 
the solid histogram represents the 
Monte Carlo simulated distribution. 
The flavor composition of the Monte Carlo distribution is shown.
\vspace{5mm}
 
\noindent
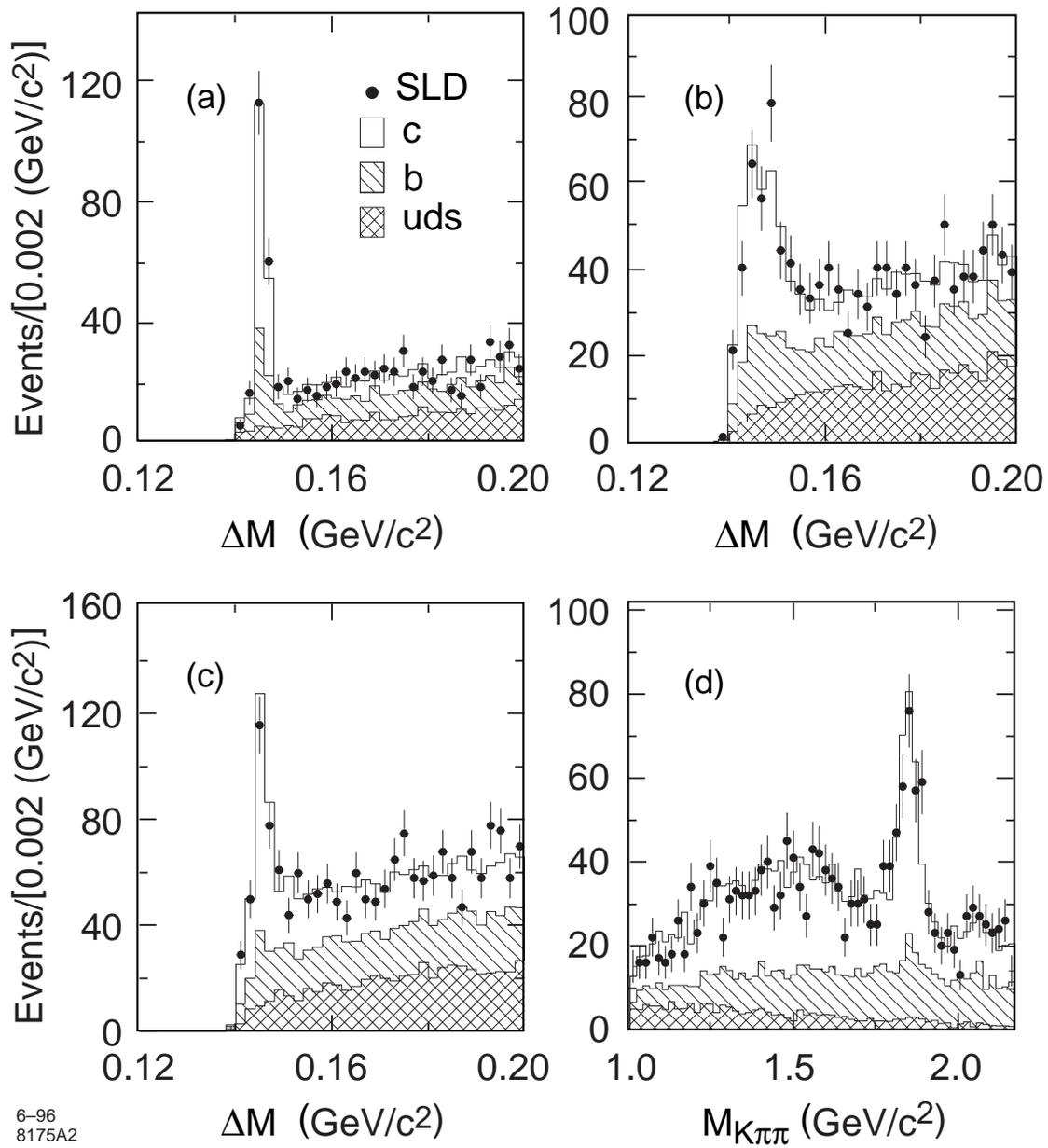
{\bf Figure 2}.
The distributions of $\Delta M$ for 
a) $D^0 \to K^-\pi^+$, b) $D^0 \to K^-\pi^+\pi^0$ and 
c) $D^0 \to K^-\pi^+\pi^+\pi^-$;
d) $M(K^-\pi^+\pi^+)$ distribution for $D^+ \to K^-\pi^+\pi^+$ (see text).
The points represent the data distributions and
the solid histograms represent the Monte Carlo simulated distributions.
The flavor composition of the Monte Carlo distributions is shown. 
\vspace{5mm}

\noindent
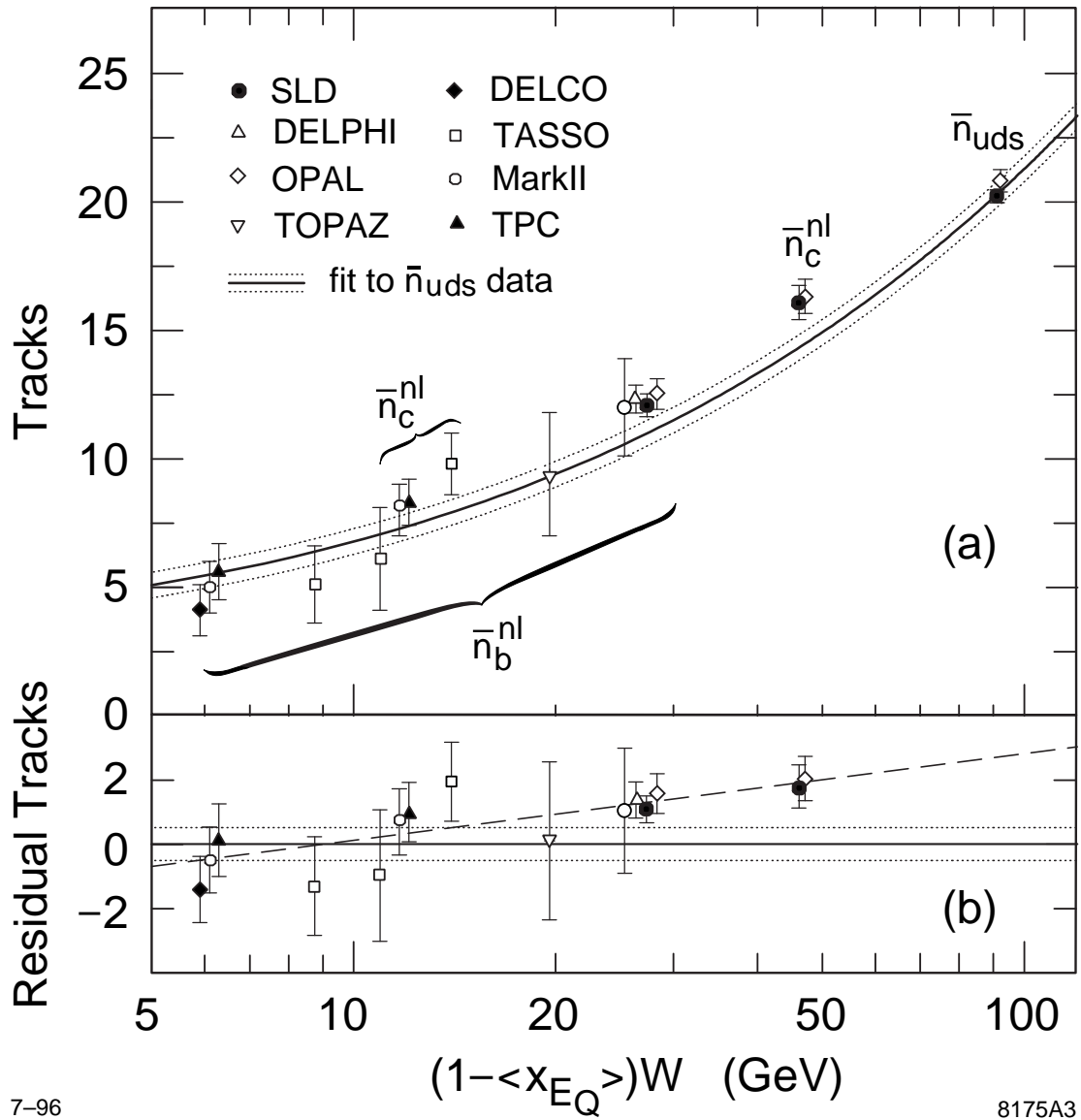
{\bf Figure 3}.
a) 
Our measurements of $\aven_{uds}$ plotted at $W=M_{Z^0}$ and the
non-leading multiplicities $\aven^{nl}_c$ and $\aven^{nl}_b$
plotted at the appropriately reduced non-leading energy
$[1-\langle x_{E_Q}(W)\rangle ]W$. 
Previous measurements of these quantities 
\cite{rowson,delco,tpc,tasso,delphi,opal,markii2,topaz}
are also shown.
The solid curve is a fit \cite{markii2}
to $\aven_{uds}$ measured in the range 5 $<W<$ 92 GeV.
The error on this curve (dotted lines) is 
dominated by the uncertainty on the removal of the heavy quark contribution
to each measured total charged multiplicity.
b) The differences (points) between the non-leading data points
in a) and the solid curve.
A linear fit to these differences is shown by the dashed line.
For clarity the different data points at the same energy are displayed
with small relative displacements in $W$.
\vspace{5mm} 

\noindent
{\bf Figure 4}.
Multiplicities differences
a) $\Delta\aven_c$ and b) $\Delta\aven_b$ 
as functions of c.m. energy.
The predictions of Ref. [3] are shown as the  
solid lines and 
those of Ref. [4] are shown as the
dashed lines.
For clarity the different data points at the same energy are displayed
with small relative displacements in $W$.
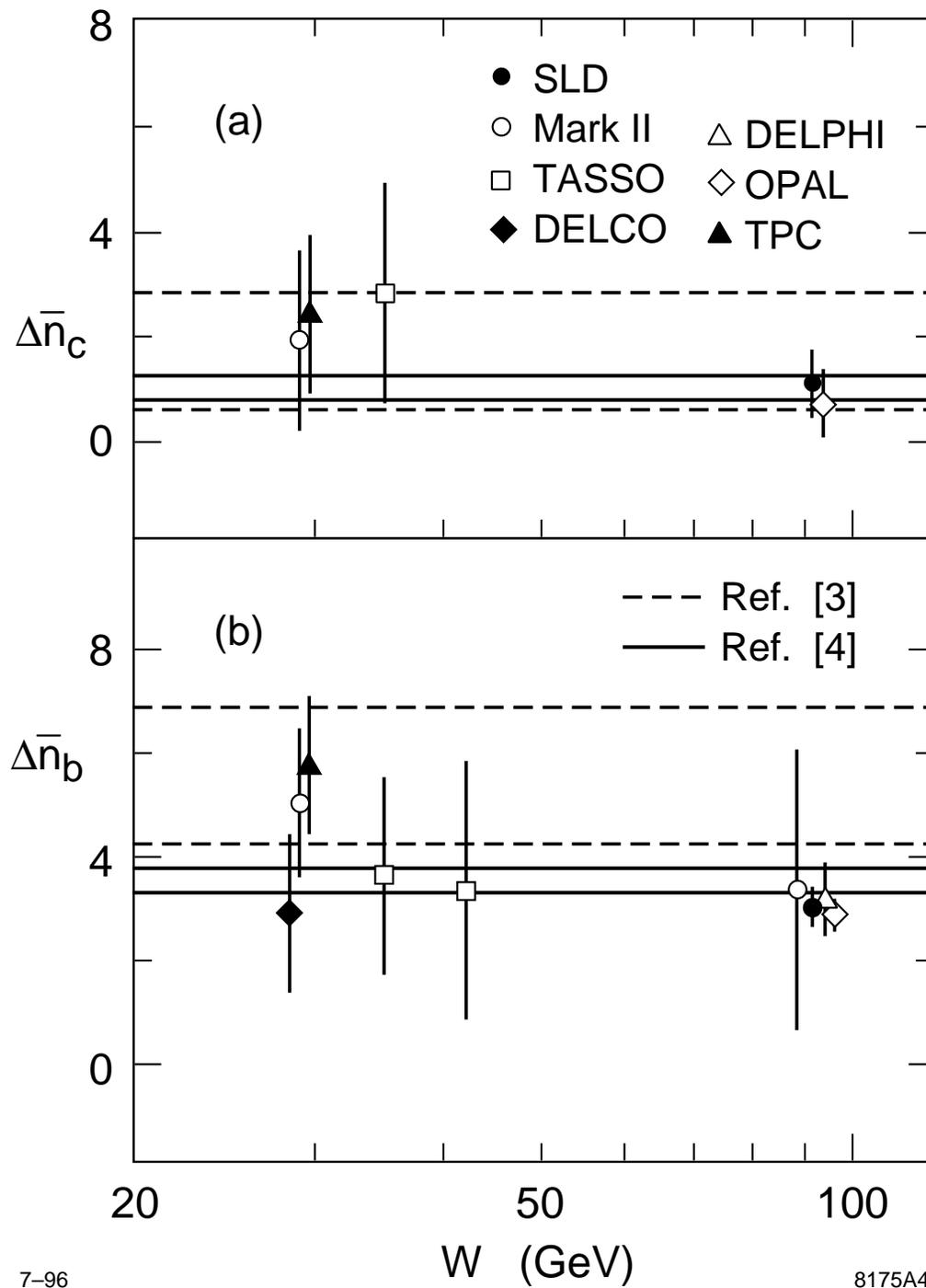
\end{document}